\documentclass[10.5pt]{article} 

\makeatletter\renewcommand{\section}{\@startsection
	{section}{1}{\z@}{-3.5ex plus -1ex minus
		-.2ex}{2.3ex plus .2ex}{\bf }}

\makeatletter\renewcommand{\subsection}{\@startsection{subsection}{2}{\z@}{-3.25ex
		plus -1ex minus
		-.2ex}{1.5ex plus .2ex}{\it }}
\makeatletter\renewcommand{\subsubsection}{\@startsection{subsubsection}{3}{-2.45ex}{-3.25ex
		plus -1ex minus -.2ex}{1.5ex plus .2ex}{\it }}

\makeatletter \@addtoreset{equation}{section}

\usepackage{graphicx}
\usepackage{float} % Kg
\usepackage{epsfig} 
\usepackage{verbatim}
\usepackage{hyperref}
\usepackage{easyfig}
\usepackage{bbm}
\usepackage{amsmath}
\usepackage{amsfonts}
\usepackage[compact]{titlesec}
\usepackage{subdepth}
\usepackage{dsfont}
\usepackage{braket}
\usepackage{tikz}
\usepackage[utf8]{inputenc}
\usepackage[T1]{fontenc}
\usepackage{amsmath}
\usepackage[toc]{glossaries}

\usepackage{caption}
\usepackage{subcaption}

\renewenvironment{thebibliography}[1]
{\baselineskip=16pt plus 2pt minus 1pt
	\section*{\large\refname
		\@mkboth{\MakeUppercase\refname}{\MakeUppercase\refname}}%
	\list{\@biblabel{\@arabic\c@enumiv}}%
	{\settowidth\labelwidth{\@biblabel{#1}}%
		\leftmargin\labelwidth
		\advance\leftmargin\labelsep
		\@openbib@code
		\usecounter{enumiv}%
		\let\p@enumiv\@empty
		\renewcommand\theenumiv{\@arabic\c@enumiv}}%
	\sloppy
	\clubpenalty4000
	\@clubpenalty \clubpenalty
	\widowpenalty4000%
	\sfcode`\.\@m}

\let\fn\footnote
\renewcommand{\footnote}[1]{\linespread{1.1}\fn{#1}\linespread{1.29}}

\usepackage{physics}
\usepackage{amsfonts}
\usepackage{amssymb}
\usepackage{mathrsfs}
\usepackage{amsmath,amssymb}
\usepackage{bm}
\usepackage{caption}
\usepackage{amsmath}
\usepackage{graphicx}
\usepackage{cite}

\usepackage[T1]{fontenc}
\usepackage[utf8]{inputenc}

\def\tyng(#1){\hbox{\tiny$\yng(#1)$}}

\newcommand{\be}{\begin{equation}}
	\newcommand{\ee}{\end{equation}}
\newcommand{\bea}{\begin{array}}
	\newcommand{\ea}{\end{array}}
\newcommand{\beqa}{\begin{eqnarray}}
	\newcommand{\eeqa}{\end{eqnarray}}

\usepackage{pdfpages}

\begin{document}
	
	%%%%%%%%%%%%%%%%%%%%%%%%%%%%%%%%%%%%%%%%%%%%%%%
%%%%%%%%%%%%%%%%%%%%%%%%%%%%%%%%%%%%%%%%%%%%%%%
%\fontfamily{pnb}\fontsize{12pt}{16pt}\selectfont
%\fontfamily{pzc}\fontsize{14pt}{16pt}\selectfont
%\fontfamily{pbk}\fontsize{12pt}{16pt}\selectfont
%\fontfamily{cmr}\fontsize{11pt}{15pt}\selectfont
\fontfamily{bch}\fontsize{11pt}{15pt}\selectfont
%\fontfamily{phv}\fontshape{ro}\fontsize{11pt}{14pt}\selectfont
%\fontfamily{ptm}\fontseries{m}\fontshape{r}\fontsize{12pt}{16pt}\selectfont
%\fontfamily{pnc}\fontseries{m}\fontshape{r}\fontsize{11pt}{15pt}\selectfont
%\fontfamily{ppl}\fontseries{m}\fontshape{r}\fontsize{11pt}{15pt}\selectfont
%\usefont{T1}{phv}{m}{it}
%%%%%%%%%%%%%%%%%%%%%%%%%%%%%%%%%%%%%%%%%%%%%%%
%%%%%%%%%%%%%%%%%%%%%%%%%%%%%%%%%%%%%%%%%%%%%%%
	\begin{titlepage}
	\begin{flushright}
		
	\end{flushright}
	
\vskip 5em
	
	\begin{center}
		{\Large \bf Chaos in the Mass-Deformed ABJM Model}\\
		~\\
	
		\vskip 3em
		
		\centerline{$ \text{\large{\bf{S. K\"{u}rk\c{c}\"{u}o\v{g}lu}}}$}
		\vskip 0.5cm
		\centerline{\sl Middle East Technical University, Department of Physics,}
		\centerline{\sl Dumlupınar Boulevard, 06800, Ankara, Turkey}
		\vskip 1em
	
		\vskip 1em
		\begin{tabular}{r l}
			E-mail: 
			&\!\!\!{\fontfamily{cmtt}\fontsize{11pt}{15pt}\selectfont kseckin@metu.edu.tr}
		\end{tabular}
		
	\end{center}
	
	\begin{quote}
		\begin{center}
			{\bf Abstract}
		\end{center}
		
	Chaotic dynamics of the mass deformed ABJM model is explored. To do so, we consider spatially uniform fields and obtain a family of reduced effective Lagrangians by tracing over ansatz configurations involving fuzzy two-spheres with collective time dependence. We examine how the largest Lyapunov exponent, $\lambda_L$, changes as a function of $E/N^2$, where $N$ is the matrix size. In particular, we inspect the temperature dependence of $\lambda_L$ and present upper bounds on the temperature above which $\lambda_L$ values comply with the MSS bound, $ \lambda_L \leq 2 \pi T $, and below which it will eventually be not obeyed.
		
	\vskip 2em
	
	To be published in Particles, Fields and Topology: Celebrating A.P. Balachandran, a Festschrift volume for A.P. Balachandran (World Scientific Publishing Co., Singapore).

	\end{quote}
	
\end{titlepage}

\setcounter{footnote}{0}
\pagestyle{plain} \setcounter{page}{2}

\section{Introduction}
\label{intro}

Research work exploring the structure of chaotic dynamics emerging from the matrix gauge theories has become quite abundant recently \cite{Sekino:2008he,  Gur-Ari:2015rcq, Maldacena:2015waa, Maldacena:2016hyu, Buividovich:2017kfk, Coskun:2018wmz, Baskan:2019qsb}. These studies are propelled by a result due Maldacena-Shenker-Stanford (MSS) \cite{Maldacena:2015waa}, which briefly states that the largest Lyapunov exponent (which is a measure of chaos in both classical and quantum mechanical systems) for quantum chaos is controlled by a temperature-dependent bound and given by $\lambda_L \leq 2\pi T$. Systems which are holographically dual to the black holes are conjectured to be maximally chaotic, i.e. saturate this bound. This is already proved for the Sachdev-Ye-Kitaev (SYK) \cite{Maldacena:2016hyu}  model, and expected to be so for other matrix models which have a holographic dual such as the BFSS \cite{Banks:1996vh} model. In \cite{Gur-Ari:2015rcq}, classical chaotic dynamics of the Banks-Fischler-Shenker-Susskind (BFSS) model \cite{Banks:1996vh}, which provides a good approximation of the quantum theory in the high temperature limit, is studied and it was found that the largest Lyapunov exponent scales as $\lambda_L = 0.2924(3) (\lambda_{'t \, Hooft} T)^{1/4}$ and therefore the MSS bound is violated only at temperatures below the critical temperature $T_c \approx 0.015$, while it remains parametrically smaller than $2 \pi T$ for $T > T_c$. In, \cite{Baskan:2019qsb} chaotic dynamics of massive deformations of the bosonic sector of the BFSS model was explored by exploiting matrix configurations involving fuzzy spheres and upper bounds on the critical temperature, $T_c$, are estimated.

In a more recent paper \cite{Baskan:2022dys}, we have examined chaos emerging from the massive deformation of the Aharony-Bergman-Jafferis-Maldacena (ABJM) model. Here, I will be reporting on a part of this work. In brief our focus can be indicated as follows. ABJM model is a supersymmetric Chern-Simons (CS) gauge theory coupled to matter fields and describes the dynamics of $N$ coincident $M2$-branes \cite{Aharony_2008, Nastase:2015wjb}. It possesses a massive deformation preserving all the supersymmetry, but breaking the $R$-symmetry. The vacuum configurations in this model are fuzzy two-spheres described in terms of Gomis, Rodriguez-Gomez, Van Raamsdonk and Verlinde (GRVV) matrices \cite{Gomis_2008}. For the purpose of studying the chaotic dynamics, we reduce this model from $2+1$ to $0+1$ dimensions by considering that the fields are spatially uniform and work in the 't Hooft limit. Tracing over an ansatz fuzzy two-sphere matrix configuration with collective time dependence, we obtain a family of effective Hamiltonians. Solving the equations of motion numerically, we examine how the largest Lyapunov exponent, $\lambda_L$, changes as a function of $E/N^2$. Making use of the virial and equipartition theorems, we investigate the implications for the aforementioned MSS conjecture. The main outcomes of our work are the upper bounds we obtain on the temperatures above which largest Lyapunov exponents comply with the MSS bound and below which it will eventually be not obeyed.

\section{Mass-Deformed ABJM Model with Spatially Uniform Fields}\label{DimenABJM}

Bosonic part of the ABJM model \cite{Aharony_2008} is an $SU(N)_k \times SU(N)_{-k}$ Chern-Simons gauge theory in $2+1$ dimensions. The subscripts $\pm k \in {\mathbb Z}$ label the level of the Chern-Simons terms associated with these gauge fields. The model involves the connections $A_{\mu}$ and $\hat{A}_{\mu}$ ($\mu:0,1,2$) transforming in the standard manner under the $SU(N)_k$ and $SU(N)_{-k}$ gauge transformations, as well as the complex scalar fields $(Q^\alpha, R^\alpha)$ which transform bi-fundamentally under the gauge symmetry, i.e. in the form $Q^\alpha \rightarrow U_L Q^\alpha U_R$,  $R^\alpha \rightarrow U_L R^\alpha U_R$, where $(U_L, U_R) \in SU(N)_k \times SU(N)_{-k} $. 

In order to dimensionally reduce $S_{ABJM}$ to $0+1$ dimensions we declare that both the gauge fields and the complex scalar are spatially uniform, i.e. independent of the spatial coordinates and depend on time only. Consequently, all partial derivatives with respect to the spatial coordinates vanish. We introduce the notation $A_\mu \equiv (A_0,X_i)$, $\hat{A}_\mu \equiv (\hat{A}_0, \hat{X}_i)$ with $(i=1,2)$. The action takes the form 
\begin{multline}
S_{ABJM-R} = N \int d t \, \, -\frac{k}{4\pi} \Tr(\epsilon^{ij} X_i {\mathcal D}_0 {X}_j) + \frac{k}{4\pi}\Tr(\epsilon^{ij} \hat{X}_i  {\hat {\mathcal D}}_0 {\hat{X}}_j ) \\ + \Tr(|D_0  Q^\alpha|^2) -  \Tr(|D_i Q^\alpha|^2)  +  \Tr(|D_0  R^\alpha|^2) -\Tr(|D_i R^\alpha|^2) - V \,,
\label{ABJMR2}
\end{multline}
where $D_i  Q^{\alpha} = iX_i Q^{\alpha}-i Q^{\alpha} \hat{X_i}$ \, $D_i R^{\alpha} = i X_i R^{\alpha} -i R^{\alpha} \hat{X_i}$ and the covariant derivatives are 
$D_0 Q^{\alpha} = \partial_{0} Q^{\alpha} +i A_0 Q^{\alpha} -i Q^{\alpha} \hat{A_{0}}$, $D_0 R^{\alpha} = \partial_{0} R^{\alpha} +i A_0 R^{\alpha} -i R^{\alpha} \hat{A_{0}}$, ${\mathcal D}_0 {X}_i = \partial_0 X_i - i \lbrack A_0 \,, X_i \rbrack$, ${\hat {\mathcal D}}_0 {\hat{X}}_i = \partial_0 \hat{X}_i - i \lbrack {\hat A}_0 \,, \hat{X}_i \rbrack$.
The potential term is given as $V=\Tr(|M^{\alpha}|^2+|N^{\alpha}|^2)$ where $M^\alpha= \mu Q^{\alpha}+\frac{2\pi}{k}(2Q^{[\alpha}Q_{\beta}^{\dagger}Q^{\beta]}+R^{\beta}R_{\beta}^{\dagger}Q^\alpha-Q^\alpha R_\beta^\dagger R^{\beta} +2Q^\beta R_{\beta}^\dagger R^\alpha -2R^\alpha R_{\beta}^\dagger Q^\beta)$ and $N^\alpha= -\mu R^{\alpha}+\frac{2\pi}{k}(2R^{[\alpha}R_{\beta}^{\dagger}R^{\beta]}+Q^{\beta}Q_{\beta}^{\dagger}R^\alpha-R^\alpha Q_\beta^\dagger Q^{\beta} +2R^\beta Q_{\beta}^\dagger Q^\alpha-2Q^\alpha Q_{\beta}^\dagger R^\beta)$. Here $\mu$ stands for the masses of the fields $(Q^\alpha, R^\alpha)$ and we have used the notation $Q^{[\alpha}Q_{\beta}^{\dagger}Q^{\beta]}=Q^\alpha Q_\beta^\dagger Q^\beta-Q^\beta Q_\beta^\dagger Q^\alpha$ and likewise for $R^\alpha $'s. 

In (\ref{ABJMR2})  it is understood that all fields depend only on time. Let us also point out that this form of the action is already written in the 't Hooft limit. The latter is defined as follows. Reducing from $2+1$ to $0+1$ dimensions, we have integrated over the two-dimensional space whose volume may be denoted, say, by $V_2$. Therefore, we may introduce $\lambda_{'t \, Hooft} := \frac{N}{V_2}$ and require that it remains finite in the limit $V_2 \rightarrow \infty $ and $N \rightarrow \infty$. In  the action, $S_{ABJM-R}$, we have scaled $\lambda_{'t \, Hooft}$ to unity. Clearly, if needed, $\lambda_{'t \, Hooft}$ may be restored back in $S_{ABJM-R}$ by making the scalings $X_i \rightarrow \lambda^{-1/2}X_i$, $\hat{X_i} \rightarrow \lambda^{-1/2} \hat{X_i}$, $A_0 \rightarrow \lambda^{-1/2}A_0$, $\hat{A_0} \rightarrow \lambda^{-1/2}\hat{A_0}$, $Q_\alpha \rightarrow \lambda^{-1/4}Q_\alpha$, $R_\alpha \rightarrow \lambda^{-1/4}R_\alpha$, $\mu \rightarrow \lambda^{-1/2}\mu$ and $t \rightarrow \lambda^{1/2} t $. It should be clear from (\ref{ABJMR2}) that $S_{ABJM-R}$ is manifestly gauge invariant under the $SU(N)_k \times SU(N)_{-k}$ gauge symmetry and the reduced CS coupling $\frac{k V_2}{4 \pi}$ is no longer level quantized. A more comprehensive discussion on the latter fact may be found in \cite{Baskan:2022dys}.

The ground states are given by configurations minimizing the potential $V$. Since the latter is positive definite, its minimum is zero and is given by the configuration $ M^\alpha=0=N^\alpha$. There are two immediate solutions to this, which are given as $(R^\alpha, Q^\alpha) = (c G^\alpha, 0)$ and $(R^\alpha,  Q^\alpha) = (0, c G^\alpha)$ where $G^\alpha$ are GRVV matrices \cite{Gomis_2008,Nastase:2015wjb} defining a fuzzy two-sphere \cite{Balachandran:2005ew} at the matrix level $N$ and $c = \sqrt{\frac{k\mu}{4\pi}}$.  Explicitly, $G^\alpha$ are given as \cite{Gomis_2008} $(G^1)_{m n} =\sqrt{m-1}\,\delta_{mn}$, $(G^2)_{mn} =\sqrt{N-m}\,\delta_{m+1 \, n}$, $(G_1^\dagger)_{mn} =\sqrt{m-1}\,\delta_{m n}$, $(G_2^\dagger)_{mn} =\sqrt{N-n}\,\delta_{n+1 \, m}$ with $m,n = 1 \,, \cdots \,, N$, and they fulfill the relation $G^{\alpha} = G^\alpha G_\beta^\dagger G^\beta-G^\beta G_\beta^\dagger G^\alpha$.

In what follows, we make the gauge choice $A_0=0 =\hat{A}_0$ and therefore have the Gauss law constraints from variations of $S_{ABJM-R}$ with respect to $A_0$ and $\hat{A}_0$. These are $\frac{k}{2\pi}[X_1,X_2] + \dot{Q}^\alpha Q_\alpha^{\dagger} -Q^\alpha \dot{Q}_\alpha^{\dagger} + \dot{R}^\alpha R_\alpha^{\dagger} -R^\alpha \dot{R}_\alpha^{\dagger}= 0$ and $-\frac{k}{2\pi}[\hat{X}_1,\hat{X}_2] - Q_\alpha^{\dagger} \dot{Q}^\alpha +\dot{Q}_\alpha^{\dagger}Q^\alpha  - R_\alpha^{\dagger} \dot{R}^\alpha +\dot{R}_\alpha^{\dagger} R^\alpha = 0$.

Hamiltonian takes the form 
\be
H = \Tr \left( \frac{1}{N} |P_Q^\alpha|^2 +  \frac{1}{N} |P_R^\alpha|^2 +N  |D_i Q^{\alpha}|^2 + N | D_i R^{\alpha}|^2 \right) + N V  \,,
\ee
where $P_Q^\alpha =\frac{\partial L}{\partial {\dot Q}^\alpha} = N {\dot Q}^{\alpha \dagger}$ and $P_R^\alpha =\frac{\partial L}{\partial {\dot R}^\alpha}= N {\dot R}^{\alpha \dagger}$ are the conjugate momenta associated with $Q_\alpha$ and $R_\alpha$ respectively. Finally, we note that the scaling transformation $(Q_\alpha \,, R_\alpha) \rightarrow  (\rho^{-1/2} \, Q_\alpha \,, \rho^{-1/2} \, R_\alpha)$, $(X_i \,, \hat{X}_i) \rightarrow  (\rho^{-1} \, X_i \,, \rho^{-1}\, \hat{X}_i)$, $t \rightarrow \rho \, t $, where $\rho$ is an arbitrary positive constant. Under this scaling, $(P_Q^\alpha\,, P_R^\alpha) \rightarrow (\rho^{-3/2} \, P_Q^\alpha \,, \rho^{-3/2} \, P_R^\alpha)$ and also that $V |_{\mu = 0} \rightarrow \rho^{-3} \, V |_{\mu =0}$ indicating that the energy scales as $E \rightarrow \rho^{-3} E $. Since the Lyapunov exponent has the dimensions of inverse time, we see that it scales as $\lambda_L \propto E^{1/3}$ in the massless limit. In what follows, we will see that this scaling of the Lyapunov exponents with energy is essentially preserved after taking the mass deformations into account.  

\section{Ansatz Configuration and the Effective Action}

\label{Ansatz1sec}

We consider the matrices $X_i = \alpha(t) \, \text{diag} ((A_i)_1,(A_i)_2,\dots,(A_i)_N)$, $\hat{X}_i  = \beta(t)\, \text{diag}((B_i)_1, \\ (B_i)_2, \dots,(B_i)_N)$, $Q_\alpha = \phi_\alpha(t) \, G_\alpha\,, R_\alpha=0$, where $(A_i)_m$, $(B_i)_m$ are constants, $i =1,2$, $m = 1,2,...,N$ and $\alpha =1,2$.  No sum over the repeated index $\alpha$ is implied. Here $\phi_\alpha(t)$, $\alpha(t)$, $\beta(t)$ are real functions of time and the Gauss law constraint equations are easily seen to be satisfied by this choice of the matrices. Evaluating the equations of motion for $\alpha(t)$ and $\beta(t)$, we find that the emerging coupled equations have only one possible real solution and that is the trivial solution given simply as $\alpha(t)= \beta(t)=0$ \cite{Baskan:2022dys}. Thus, setting $X_i$ and $\hat{X_i}$ to zero from now on and  performing the traces over the GRVV matrices, we find the reduced Hamiltonians
\be 
\label{hamiltonian1}
H_N(\phi_1,\phi_2\, p_{\phi_1},p_{\phi_2}) = \frac{p_{\phi_1}^2}{2N^2(N-1)}+\frac{p_{\phi_2}^2}{2N^2(N-1)}+V_N(\phi_1,\phi_2) 
\ee
where $V_N(\phi_1,\phi_2) = N^2(N-1)(\frac{1}{2}\mu^2(\phi_1^2+\phi_2^2)+\frac{8\pi\mu}{k}\phi_1^2\phi_2^2+\frac{8\pi^2}{k^2}\phi_1^4\phi_2^2+\frac{8\pi^2}{k^2}\phi_2^4\phi_1^2)$. In the limit $\mu \rightarrow 0$, we have $H_N \rightarrow \rho^{-3} H_N$ under the scaling $(\phi_1 \,, \phi_2) \rightarrow (\rho^{-1/2} \, \phi_1 \,, \rho^{-1/2} \, \phi_2)$ and $t \rightarrow \rho \, t $, in view of the scaling properties of the matrix model given in the previous section. In what follows, we will explore the dynamics emerging from the Hamilton's equations at $\mu =1$ at several different matrix levels $N$ and the CS coupling $k$.

\subsection{Chaotic dynamics and the Lyapunov exponents}

\label{lyap1}

We examine the chaotic dynamics of the models governed by the Hamiltonians $H_N$. For this purpose we numerically evaluate the Lyapunov exponents of these models by solving the associated Hamilton's equations of motion.  As it is well known (see \cite{Baskan:2022dys} and the references therein) the largest Lyapunov exponent is essentially a measure of the sensitivity of a system to given initial conditions. More precisely, it gives the exponential growth in perturbations and in this regard it provides a quantitative means of detecting and examining chaos. The phase spaces for $H_N$ are all $4$-dimensional and their chaotic dynamics is governed by the largest (and only) positive Lyapunov exponent. Obtaining the solutions of the Hamilton's equations with $40$ ($100$) randomly selected initial conditions for $k > 1$ ($k < - 1 $) at a given energy value $E$ and matrix level $N$, we calculate the mean of the time series for each and every Lyapunov exponent from all runs. In the simulation, we take a time step of $0.25$ and run the code from time $0$ to $3000$. Results for the largest Lyapunov exponent, $\lambda_L$, as a function of $E/N^2$ for $N = 15, 25$ at several different values of the energy and $k=\pm 1, \pm2$ are discussed in the following.

\begin{figure}[!htb]
\begin{subfigure}[!htb]{0.5\textwidth}
\includegraphics[width=6.6cm]{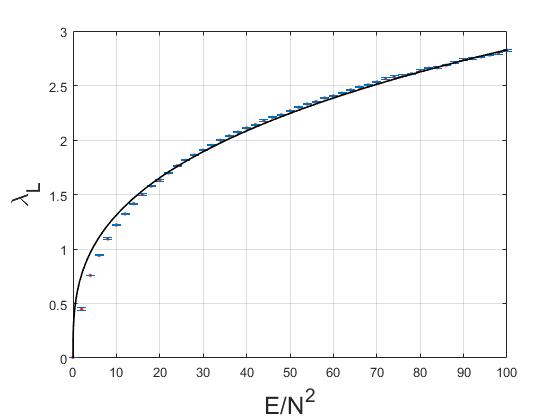}
	\caption{$k=1 \,, N=15$}
	\label{fig:len151}
\end{subfigure}
	\begin{subfigure}[!htb]{0.5\textwidth}
	\includegraphics[width=6.6cm]{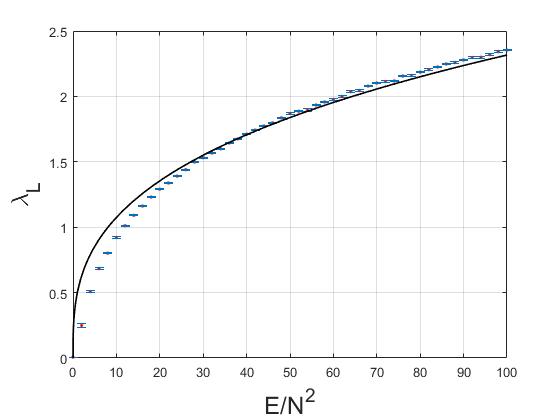}
	\caption{$k=1 \,, N=25$}
	\label{fig:len25}
\end{subfigure}
\begin{subfigure}[!htb]{0.5\textwidth}
	\includegraphics[width=6.6cm]{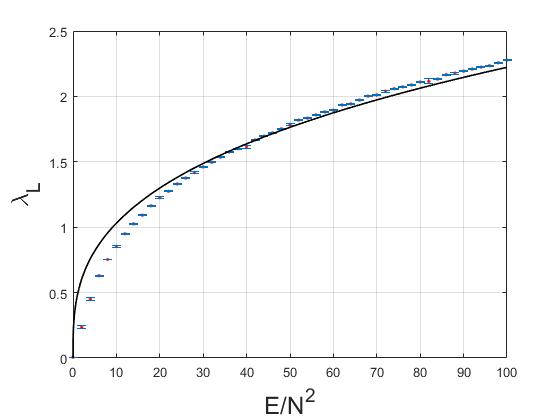}
	\caption{$k=2 \,, N=15$}
	\label{fig:len15k2}
\end{subfigure}
\begin{subfigure}[!htb]{0.5\textwidth}
	\includegraphics[width=6.6cm]{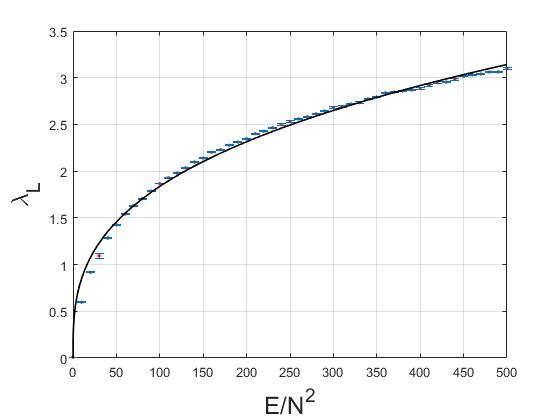}
	\caption{$k=2 \,, N=25$}
	\label{fig:len25k2}
\end{subfigure}
\caption{Largest Lyapunov exponent and the best fitting curves in the form $\lambda_L = \alpha_N(\frac{E}{N^2})^{1/3}$.}
\label{figk1}
\end{figure}

For $k =1, 2$, the data and best fitting curves of the form $\lambda_L = \alpha_N(\frac{E}{N^2})^{1/3}$ are given in Figure \eqref{figk1} at $N = 15, 25$, and $k=1,2$. They clearly demonstrate the $\lambda_L \propto E^{1/3}$ dependence of the Lyapunov exponent anticipated by the scaling argument.
Values of the coefficients $\alpha_N$ for the fitting curves in Figures \eqref{figk1} are $\alpha_{15} =0.6092, 0.4788 $ and $\alpha_{25} = 0.499, 0.3958$ for $k=1,2$, respectively.

\begin{figure}[!htb]
\begin{subfigure}[!htb]{0.5\textwidth}
		\includegraphics[width=6.6cm]{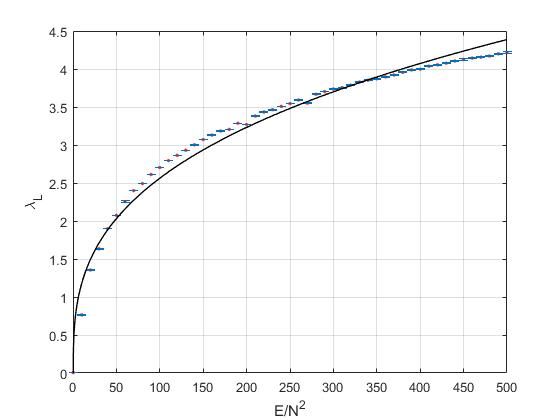}
		\caption{$k = - 1 \,,N=15$}
		\label{fig:len15k-1}
\end{subfigure}
\begin{subfigure}[!htb]{0.5\textwidth}
			\includegraphics[width=6.6cm]{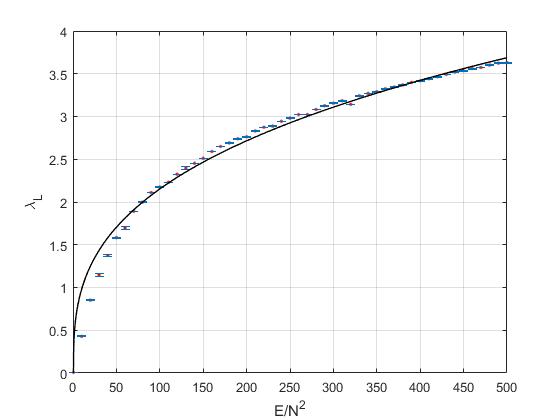}
			\caption{$k =-1 \,,N=25$}
			\label{fig:len25k-1}
\end{subfigure}
\begin{subfigure}[!htb]{0.5\textwidth}
	\includegraphics[width=6.6cm]{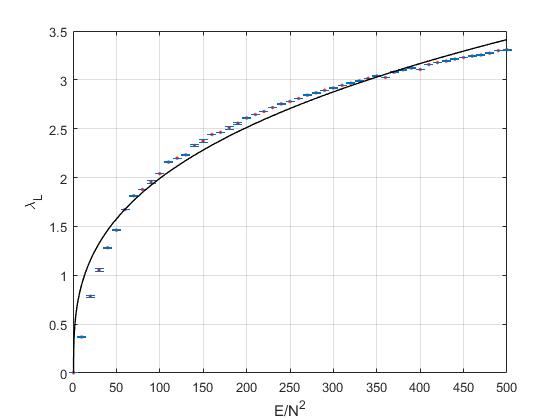}
	\caption{$k=-2 \,,N =15$}
	\label{fig:len15k-2}
\end{subfigure}
\begin{subfigure}[!htb]{0.5\textwidth}
	\includegraphics[width=6.6cm]{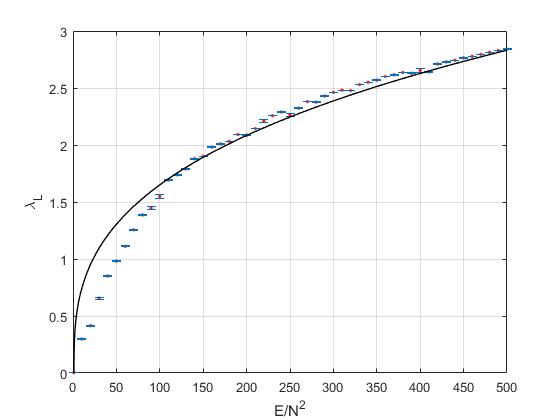}
	\caption{$k=-2 \,,N=25$}
	\label{fig:len25k-2}
\end{subfigure}

\caption{Largest Lyapunov exponent and the best fitting curves in the form $\lambda_L=\alpha_N(\frac{E}{N^2}-\gamma_N)^{1/3}$ at $k=-1,-2$.}
\label{fig-1}
\end{figure}

For $k = -1, -2$, we seek best fitting curves of the form $ \lambda_L \propto (\frac{E}{N^2}- \gamma_N)^{1/3}$ to the Lyapunov data, where $\gamma_N$ is determined by $N$, $k$ and $\mu$ and proportional to the minimum value of $\tilde{V}_N$ involving solely the quartic terms of $V_N$. This specific form of the $\gamma_N$ is indeed motivated by the virial and equipartition theorems as will be shortly discussed in the following section. Data and the fitting curves are provided in Figure \eqref{fig-1}.
The coefficients $\alpha_N$ for the fitting curves are $\alpha_{15} = 0.5529, 0.4281 $ and $\alpha_{25} = 0.4648, 0.357$ for $k=-1,-2$, respectively.

\subsection{Temperature dependence of the Lyapunov exponent} \label{Tdependence}

In the massless limit, $\lambda_{t' Hooft}$ and the temperature are the only dimensionful parameters and using dimensional analysis, we may easily see that $\lambda_L \propto (\lambda_{t' Hooft}T)^{1/3}$. This is because, in $0+1$ dimensions $\lambda_{' t Hooft}=\frac{N}{V_2}$, $V_2$ being the volume of the $2$-dimensional space, we have integrated over in going from $2+1$ to $0+1$ dimensions and it has the dimension $[Length]^{-2}$ and hence $\lambda_L$ has the dimension of $[Length]^{-1}$. In view of the equipartition theorem, this is consistent with $\lambda_{L} \propto E^{1/3}$ which was independently already noted based on the scaling symmetry. Shortly, we will also discuss the effects of mass parameter on the relation between the energy and temperature upon the application of the virial and the equipartition theorems. We may contrast these features with those of the BFSS model, in which $\lambda_L \propto (\lambda_{t' Hooft}T)^{1/4}$. Since the potential is purely quartic in this latter case, the system has a scaling symmetry implying that $\lambda_L \propto E^{1/4}$. Mass deformations in this model were examined via a method involving fuzzy spheres and similar to our present approach in \cite{Baskan:2019qsb}. 

In the model described by the ansatz configuration (after setting $X_i$ and $\hat{X_i}$ to zero) total number of real degrees of freedom is $4 N^2$ before taking the global gauge symmetry and the Gauss law constraints into account. The latter imposes only $N^2$ real relations as the two equations comprising it reduce to the same equation upon integrating by parts and taking the Hermitian conjugate of one or the other, while the $R$-symmetry gives eight real relations among the unconstrained degrees of freedom, therefore we have $n_{d.o.f.}\approx 3 N^2$ at large $N$.

Applying the virial theorem to \eqref{hamiltonian1} gives $2 \left \langle K\right \rangle = 2\left \langle V_N \right \rangle +\tilde{V}_N(\phi_1,\phi_2)$, where $\tilde{V}_N(\phi_1,\phi_2) \\ = 2N^2(N-1)\left(\frac{8\pi\mu}{k} \phi_1^2\phi_2^2 +\frac{16\pi^2}{k^2}\phi_1^2\phi_2^4 +\frac{16\pi^2}{k^2}\phi_1^4\phi_2^2 \right)$. The latter is positive definite if $k$ and $\mu$ have the same sign, but for $k$ and $\mu$ with opposite signs we have $Min (\tilde{V}_N(\phi_1,\phi_2) ) = N^2(N-1) \frac{4 k\mu^3}{27\pi}$. Applying the equipartition theorem to the kinetic energy yields $\left \langle K \right \rangle \approx \frac{3}{2} N^2 T$ at large $N$. In what follows, we consider $\mu =1$.

For $k \geq1$, we have the inequality $\left \langle K \right \rangle \geq \left \langle V_N \right \rangle$. Together with the results of the equipartition theorem this implies that $\langle E \rangle = \left \langle K \right \rangle + \left \langle V_N \right \rangle \leq n_{d.o.f} T \approx 3N^2 T$. We can express this inequality in the form $\frac{E}{N^2} \leq 3 T$, where we have also dropped the brackets on energy for ease in notation. We may compare and relate this result to the the MSS bound $\lambda_L \leq 2\pi T$ on the largest Lyapunov exponent for quantum chaos \cite{Maldacena:2015waa}. ABJM model has a gravity dual \cite{Nastase:2015wjb} via the AdS/CFT correspondence and we may expect the MSS conjecture to hold for quantum chaotic dynamics of the ABJM model too. Since our analysis here is confined to the classical regime, we should expect that the MSS bound be eventually not obeyed at sufficiently low temperatures. Indeed, from our results, we observe that there is a critical temperature, which we may denote as $T_c$ and given by solving the equation $\alpha_N(3T)^{1/3}=2\pi T$. This yields $T_c = \sqrt{3}\left(\frac{\alpha_N}{2\pi}\right)^{3/2}$. This is an upper bound for the critical temperature at or below which MSS bound will eventually be not obeyed by our model. For $N=15$, $k=1,2$ our estimates are $T_c = 0.0523, 0.0364$, respectively, while, for $N=25$, we find $T_c = 0.0388, 0.0274$ for $k=1,2$. More comprehensive results, presented in our paper, \cite{Baskan:2022dys} indicate that $T_c$ values tend to decrease with increasing matrix size. This is in agreement with the fact that 't Hooft limit is better emulated with increasing values of $N$.

For $k\leq-1$, we proceed as follows. We may write $2 \langle K \rangle = 2 \langle V_N\rangle+ \tilde{V}_N(\phi_1,\phi_2)+\abs{Min(\tilde{V}_N)} -\abs{Min(\tilde{V}_N)}$ and this implies that $\langle K \rangle\geq\langle V_N \rangle-\frac{1}{2}\abs{Min(\tilde{V}_N)}$. Therefore, we have $E  - \frac{1}{2}\abs{Min(\tilde{V}_N)} = \langle K \rangle+\langle V_N \rangle -\frac{1}{2}\abs{Min(\tilde{V}_N)} \leq n_{d.o.f.} T$. Since $\langle K \rangle \approx \frac{3}{2} N^2 T$ at large $N$, this leads to the inequality $\frac{E}{N^2} - \gamma_N \leq  3T$, where $\gamma_N : = \frac{\abs{Min(\tilde{V}_N)}}{2N^2}$. Hence, we now clearly observe the line of reasoning that led us in the previous section to consider the best fitting curves of the form $\lambda_ N = \alpha_N \left(\frac{E}{N^2} - \gamma_N\right)^{1/3}$. These curves are already given in Figure \eqref{fig-1} and they clearly represent the variation of the largest Lyapunov exponent with respect to $E/N^2$ quite well. Finally, by the same line of reasoning used earlier, we find that the critical temperatures are found to be $T_c=\sqrt{3}(\frac{\alpha_N}{2\pi})^{3/2}$.Our numerical estimates of the critical temperature are $T_c = 0.0452, 0.0308$ for $N=15$, $k=-1,-2$, respectively, and $T_c = 0.0348, 0.0234$ for $N=25$ and $k=-1,-2$.

\section{Conclusions and Outlook}

Here, we have reported on a part of our recent work \cite{Baskan:2022dys}. Our main objective was to examine the structure of chaotic dynamics emerging from the massive deformation of the ABJM model. We have approached this problem by considering an ansatz configuration involving fuzzy spheres with collective time dependence and obtained a family of effective actions parametrized by the matrix level $N$. We computed the largest Lyapunov exponent and presented its variation with respect to $E/N^2$ and demonstrated that $\lambda_L \propto (E/N^2)^{1/3}$ or $\lambda_L \propto (E/N^2 - \gamma_N)^{1/3}$ depending on the sign of $k$. This allowed us to inspect the extent the largest Lyapunov exponent complies with the MSS bound upon the use of the virial and equipartition theorems.

\section{Reminiscences and a Tribute}

I met Bal in the late summer of 1999 at Syracuse. My recollection is that he was the first person in the States to ask me about relief efforts and the situation in the aftermath of the large earthquake that took place near Propontis, the sea of Marmara, in Turkey at around the time. I joined his research group in the following year. At first, it was kind of discombobulating to work with Bal, trying to learn so many different concepts and techniques so fast, to adjust to the flow and exposure to the ideas in many diverse directions that were being discussed and debated with other students and collaborators and the frequent visitors, who were, at the time, Denjoe (O'Connor), Peter (Presnajder), and Giorgio (Immirzi). These were the days, when the new directions in non-commutative geometry and fuzzy field theories were the main theme of interest in Bal's group. 

After a while, with the persistent but always positive attitude of Bal in approaching his students, I found myself able to take up the challenge, to learn a particular problem pretty rapidly and contribute to the development and solution of research problems leading to novel publishable results. My scientific training with Bal, especially via discussions in room 316 in the physics department, which were already a classic before I arrived at Syracuse, allowed me to gain the confidence and assertiveness which helped me to build my own career path. Our close contact continued after my graduation and resulted in several papers and a book on fuzzy physics coauthored together with Sachin Vaidya. My collaboration with Bal still continues today and we are entertaining ideas on how to exploit a coproduct we have found long time ago for fuzzy spheres, may be used to model entanglement. 

Meetings in room 316, at dinner tables or pot luck parties at Bal's house often expanded into conversations on literature, art house movies and music but most assuredly into heated debates in world politics. Working with Bal also led me to travel to diverse locations all over the world, to visit the pyramids of Sun and Moon in the Mesoamerican city of Teotihuacan, to the halls of the Prado museum in Madrid and perhaps most exhilaratingly, as I still vividly recall after so many years, to being exposed to mock charges of elephants in the monsoon forests on our way to the city of Mysore.

Wishing Bal a very happy 85th birthday with the final verses of the poem ``Plea'' by Nazım Hikmet:

\begin{quote}
	{\it 
Do away with the enslaving of man by man,\\
This plea is ours,\\
To live! Like a tree alone and free, \\
Like a forest in brotherhood, \\
This yearning is ours. }
\end{quote} 

\section*{Acknowledgements}
 
I would like to thank the editors for the kind invitation to contribute to this volume. This work is supported by TÜBİTAK under the project number 118F100.

\end{document}